\documentclass{article}
\usepackage{graphicx}
\begin{document}
\title{Magnetic field stabilization for high-accuracy mass measurements on exotic nuclides}
\author{
M. Marie-Jeanne$^{1}$\footnote{Present address: CERN, Physics Department, 1211 Geneva 23, Switzerland},
J. Alonso$^{3}$,
K. Blaum$^{3,4}$, 
S. Djekic$^3$,\\
M. Dworschak$^4$,
U. Hager$^5$,
A. Herlert$^2$\footnote{Corresponding author, email: alexander.herlert@cern.ch},
Sz. Nagy$^3$,\\
R. Savreux$^4$,
L. Schweikhard$^6$,
S. Stahl$^7$,
C. Yazidjian$^4$\\
\small{$^1$Universit\'e de Caen Basse-Normandie, 14032 Caen Cedex, France}\\
\small{$^2$CERN, Physics Department, 1211 Geneva 23, Switzerland}\\
\small{$^3$Johannes Gutenberg-Universit\"at, Institut f\"ur Physik, 55099 Mainz, Germany}\\
\small{$^4$GSI, Planckstr. 1, 64291 Darmstadt, Germany}\\
\small{$^5$University of Jyv\"askyl\"a, Department of Physics, P.O. Box 35 (YFL), 40014 Jyv\"askyl\"a, Finland}\\
\small{$^6$Ernst-Moritz-Arndt-Universit\"at, Institut f\"ur Physik, 17487 Greifswald, Germany}\\
\small{$^7$Elektronik-Beratung Dr. Stefan Stahl, Kellerweg 23, 67582 Mettenheim, Germany}\\
}
\date{\today\\ 
{\vspace{5mm}submitted to NIM\,A}}
\maketitle
\begin{abstract}
The magnetic-field stability of a mass spectrometer plays a crucial role in precision
mass measurements. In the case of mass determination of short-lived nuclides
with a Penning trap, major causes of
instabilities are temperature fluctuations in the vicinity of the trap and
pressure fluctuations in the liquid helium cryostat of the superconducting magnet.
Thus systems for the temperature and pressure stabilization of
the Penning trap mass spectrometer ISOLTRAP at the ISOLDE facility at CERN
have been installed. A reduction of the fluctuations by at least one order of magnitude
downto $\Delta T\approx\pm5$\,mK and $\Delta p\approx50$\,mtorr
has been achieved, which corresponds to a relative frequency change of $2.7\times10^{-9}$ and
$1.5\times10^{-10}$, respectively. With this stabilization the frequency determination
with the Penning trap only shows a linear temporal drift over several hours
on the 10 ppb level due to the finite resistance of the superconducting magnet coils.
\end{abstract}

\section{Introduction}
The exact masses of stable and exotic nuclides are important input values
for many areas of physics. Since the mass defect is intrinsically
connected to the binding energy of the nucleus, the nuclear properties are reflected
by the nuclear masses and mass differences, e.g. the $Q$-values of nuclear decays.
Mass values can therefore contribute to tests of nuclear models and nuclear structure
\cite{Lunney}, where especially nuclides far from the valley of stability are
of interest. Low production rates and half-lives of only a few tens of milliseconds
make high-accuracy mass measurements for these exotic nuclides a challenge.   

Depending on the respective application of mass values, the required relative mass
uncertainty ranges from $10^{-6}$ in nuclear physics to $10^{-11}$ in atomic physics
and metrology \cite{Blaum2006,SchweikIJMS251}. One example is the precise and accurate
determination of atomic masses for nuclides that exhibit a superallowed beta
decay, where besides the half-life, the branching ratio, and theoretical corrections,
the $Q$-value and thus the mass contributes to a test of the unitarity of the
Cabibbo-Kobayashi-Maskawa quark-mixing matrix of the Standard
Model \cite{HardyTowner,KellerbauerPRL,Mukherjee,BollenCalcium,SavardMg,HagerGa}.

Mass measurements at this level of precision are possible with Penning traps
\cite{Bollen}, where the cyclotron frequency
\begin{equation}
\label{eq:CyclotronFreq}
\nu_c=\frac{1}{2\pi}\frac{qB}{m}
\end{equation}
of the stored ions with mass $m$ and charge $q$ in a strong
homogeneous magnetic field $B$ can be monitored, e.g, with a
time-of-flight (TOF) detection technique \cite{Graeff}. 

The magnetic field amplitude $B$ is calibrated by measuring the
cyclotron frequency of a stable nuclide with well-known mass.
Usually the magnetic
field is sampled before and after the frequency measurement
of the nuclide of interest, and the field amplitude is linearly
interpolated between the two reference measurements \cite{KellerbauerEPJD}.
Any fluctuation of the field can thus lead to a deviation
of the deduced cyclotron frequency from the actual value and therefore
to the loss of accuracy. Especially
in the case of exotic nuclides with low production yields, for which
a frequency measurement can last up to several tens of minutes
or even a few hours, a stable magnetic field is crucial. 

The amplitude of the magnetic field of a superconducting magnet is influenced
by internal and external environmental parameters and thus, if these
are not constant, the $B$-field may vary as a function of time. For the superconducting
magnets of the ISOLTRAP experiment \cite{ISOLTRAPsetup,ISOLTRAPsetup2} there is a decay
of the field strength due to the finite resistance of the superconducting
coils (flux creep \cite{Anderson64}). In addition, temperature fluctuations in the room temperature bore
of the magnet lead to fluctuations of the magnetic field amplitude, since the
magnetic susceptibility of the material inside the $B$-field changes as a
function of temperature. Furthermore, pressure fluctuations in the helium recovery
line also influence the magnetic field strength, because the temperature of the
liquid helium, in which the superconducting coils are located, depends on the pressure
in the cryostat of the magnet.

Recently, regulations of both the helium reservoir pressure and the temperature
in the warm bore of the superconducting magnet have been installed
and tested at the SMILETRAP Penning trap mass spectrometer for highly-charged
stable ions \cite{SMILETRAP}.
In the present work, a similar implementation of both
regulation systems at the ISOLTRAP on-line mass spectrometer for short-lived ions
is described and the reduction of fluctuations is specified with respect to future
improvements in the mass determination of exotic nuclides.

\section{Experimental setup and frequency determination}
\begin{figure}
\begin{center}
\includegraphics[scale=1.0]{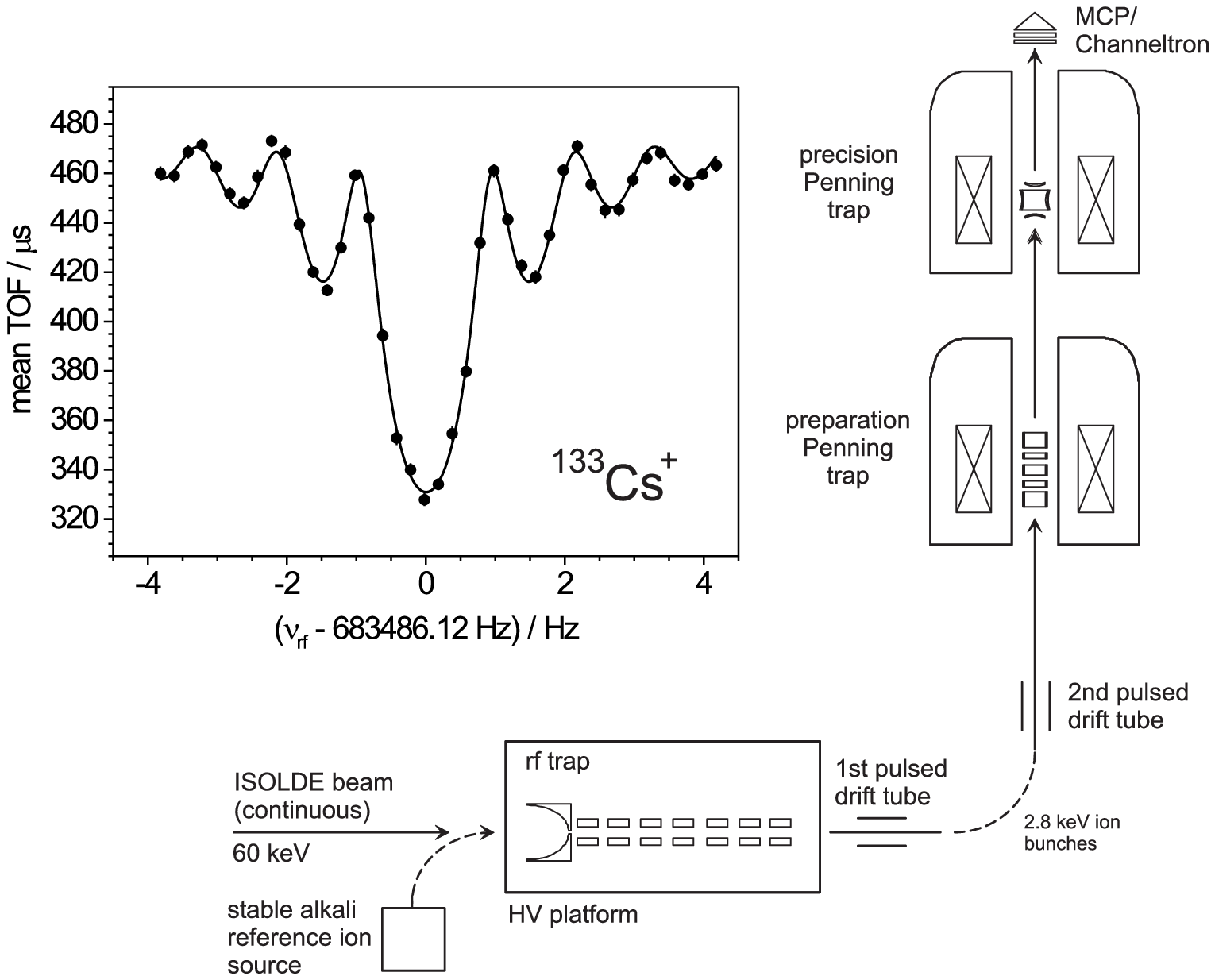}
\end{center}
\caption{\label{fig:1}
Sketch of the experimental setup of ISOLTRAP. The inset shows a cyclotron resonance
of $^{133}$Cs$^+$. The error bars of the data points are smaller than the symbol size.}
\end{figure}

For the preparation and high-accuracy mass measurement of radionuclides a combination of
three ion traps is used at ISOLTRAP (see Fig.\,\ref{fig:1}). The experimental
setup has been described in detail in \cite{ISOLTRAPsetup,ISOLTRAPsetup2} and here
only a brief summary is given.

The first ion trap is a radiofrequency quadrupole (RFQ) structure in a helium
buffer-gas environment \cite{HerfurthRFQ}, which stops, cools, and bunches the
continuous 60-keV radioactive ion beam from the target/ion-source system of
ISOLDE \cite{KuglerISOLDE}.
The ion bunch is transferred to the preparation Penning trap
\cite{PrepTrap} for further cooling and removal of contaminating isobaric ions 
with a buffer-gas cooling technique \cite{Savard}. The mass-selected
ensemble of ions is finally transferred to the precision Penning trap for
mass measurement with the TOF cyclotron-resonance detection technique \cite{Graeff}. 

The measurement principle includes first a dipolar radiofrequency (rf) excitation of the 
low-frequency magnetron motion and second a quadrupolar rf excitation.
If the quadrupolar excitation is in resonance with the cyclotron frequency,
i.e. $\nu_{rf}=\nu_c$, the magnetron motion is converted into the
high-frequency cyclotron motion of the same radius and thus the ions gain
maximum radial kinetic energy \cite{Bollen1990,Koenig}. The ions are ejected from
the trap and slowly drift upstream towards an ion detector. Along the way the ions
drift through a magnetic field gradient where the initial radial energy is converted into axial
kinetic energy due to the coupling of the magnetic moment to the field gradient \cite{Graeff}.
Thus resonantly excited ions experience a larger acceleration than non-excited
ones and therefore reach earlier the detector. 

By scanning the excitation frequency around
the expected cyclotron frequency, a TOF cyclotron resonance curve is obtained, as shown for
$^{133}$Cs$^+$ in the inset of Fig.\,\ref{fig:1}. Due to the finite excitation duration
the resonance curve exhibits a characteristic shape which is related to the Fourier transform.
The shape of the curve is well known \cite{Koenig} and can be fitted to the data points (solid line).
For the investigation of the temperature and pressure dependence, the cyclotron
frequencies $\nu_c$ of $^{85}$Rb$^+$ and $^{133}$Cs$^+$ ions from the alkali reference ion
source of ISOLTRAP have been monitored as a function of time. 

\section{Magnetic field stabilization}

\subsection{Temperature-frequency correlation}
\begin{figure}
\begin{center}
\vspace{11cm}
\includegraphics[scale=0.8]{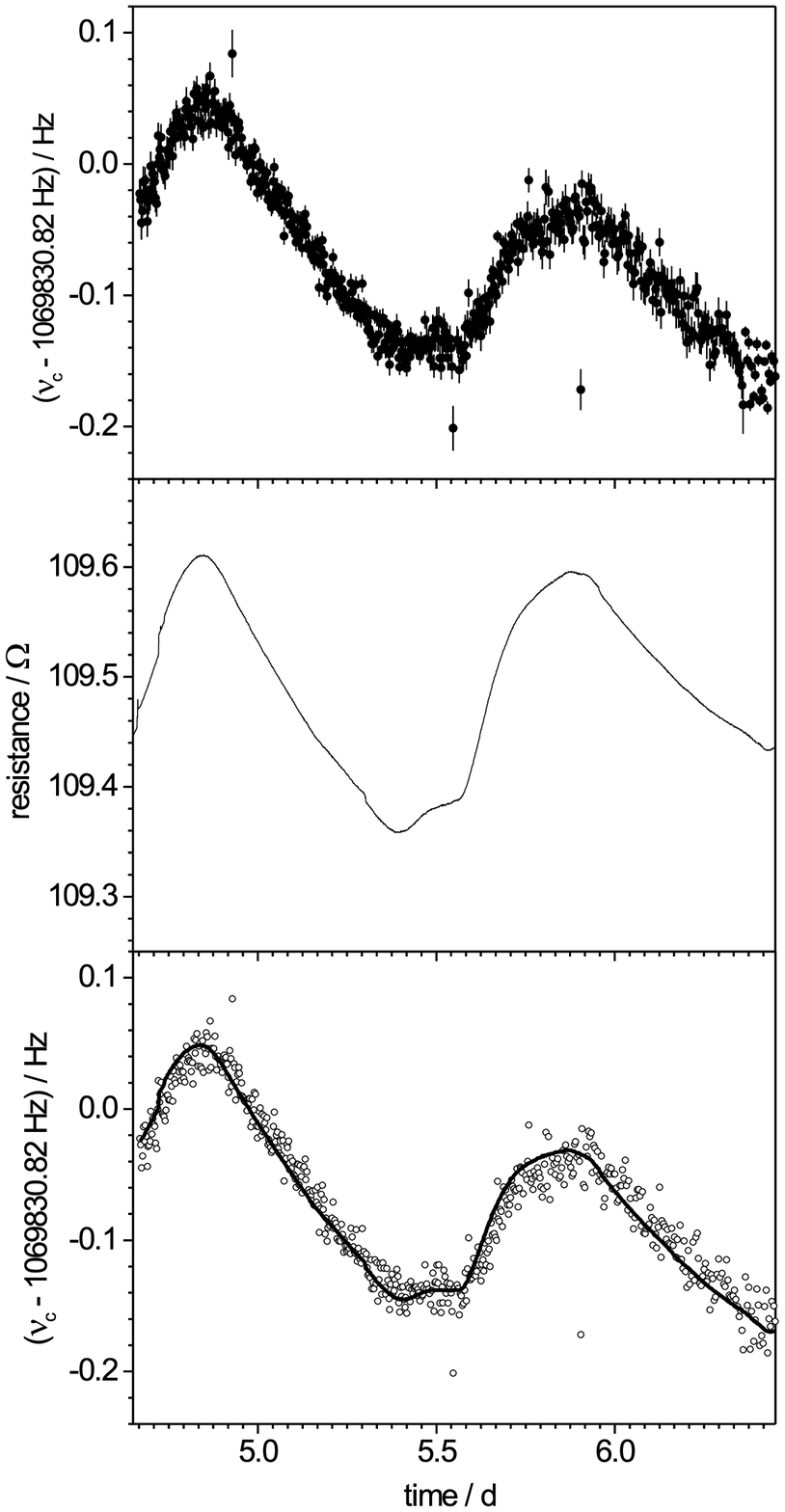}
\end{center}
\vspace{1cm}
\caption{\label{fig:2}
Top: Cyclotron frequency of $^{85}$Rb$^+$ as a function of time (data from \cite{BlaumJPhysG}).
Center: Resistance of a Pt100 sensor mounted in the vicinity of the
Penning trap vacuum tube.
Bottom: Expected behavior (solid line) of the cyclotron frequency (open circles, same data as in the top
graph) as deduced from a $\chi^2$-minimization of Eq.\,(\ref{Eq:resistance}).}
\end{figure}

The correlation between the temperature in the warm bore of the
superconducting magnet and the cyclotron frequency of ions
stored in the precision Penning trap has been observed recently at ISOLTRAP
\cite{BlaumJPhysG}. The temperature change was monitored with the
change of the resistance of a Pt100 sensor (platinum resistance thermometer)
mounted in the vicinity
of the Penning trap vacuum tube. The cyclotron frequency
of $^{85}$Rb$^+$ and the measured resistance are plotted as a function
of time in Fig.\,\ref{fig:2}. The variation
of the cyclotron frequency and of the resistance, and thus of the temperature, shows
a strong correlation. If the resistance data points are scaled
relative to the frequency data and if a linear temporal
drift of the magnetic field strength is taken into account, i.e. a linear
drift of the cyclotron frequency, the two curves nicely match
as shown in Fig.\,\ref{fig:2}\,(bottom). The frequency
can be described by 
\begin{equation}
\label{Eq:resistance}
\nu_c(R,t)=a(R-R_0)-bt+c,
\end{equation}
where $R_0=109.5\,\Omega$, $a=0.617(5)$\,Hz$\Omega^{-1}$,
$b=0.0685(6)$\,Hzd$^{-1}$, and $c=1069831.132(3)$\,Hz.
The linear field drift is thus of the order of
\begin{equation}
\frac{1}{B}\frac{dB}{dt}=-\frac{b}{c}=(-2.67\pm0.02)\times10^{-9}\,\mbox{h}^{-1}.
\end{equation}
Note that the previously reported linear drift of the magnetic field of ISOLTRAP
with a value of $(-2.30\pm0.03)\times10^{-8}$\,h$^{-1}$ \cite{KellerbauerEPJD}
has most likely a wrong exponent.

\subsection{Temperature stabilization}
Although the ISOLDE experimental hall is equipped with an air condition,
large temperature fluctuations of about 0.5\,K are observed during a day.
At ISOLTRAP these lead to a change of the cyclotron
frequency of the order of e.g. 0.1\,Hz for $^{133}$Cs$^+$ or a relative frequency
change of almost 100\,ppb.
Therefore, a temperature regulation is advantageous for the investigation of
radionuclides with a low production yield, for which long measurement periods of
more than one hour are needed to record a sufficiently large number of ions to
obtain a cyclotron resonance.

\begin{figure}
\begin{center}
\includegraphics[scale=1.3]{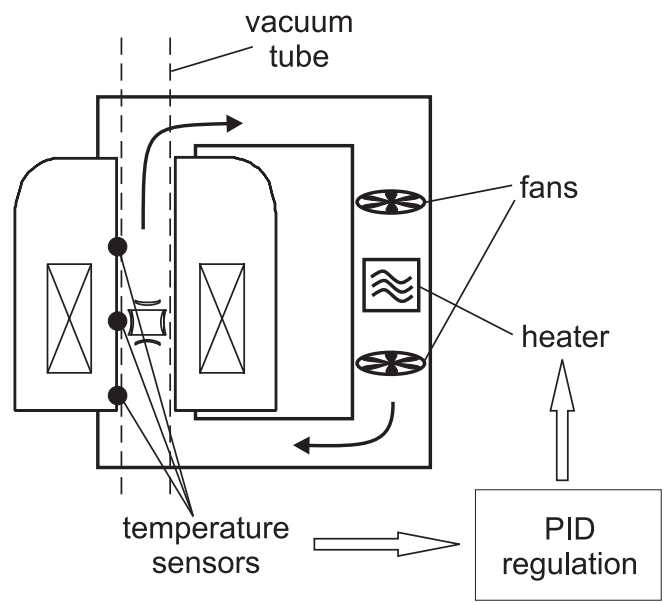}
\end{center}
\caption{\label{fig:3}
Layout of the temperature regulation system. The middle temperature
sensor is used for the PID regulation. The other two monitor the heat flow
through the warm bore of the magnet.}
\end{figure}

The layout of the temperature stabilization system is shown
in Fig.\,\ref{fig:3}. A closed aluminum tube is attached to both sides
of the room temperature bore of the magnet.
The tube encloses a heater,
two fans, and three temperature sensors. These 
integrated circuit transducers (AD590 from Analog Device) 
produce an output current proportional to the absolute temperature. 
For a supply voltage in the range between 4\,V and 30\,V each sensor
acts as a high-impedance constant-current regulator with 1\,$\mu$A/K.
The devices are calibrated to $298.2\,\mu$A output at
a temperature of $298.2$\,K. In the present
system a flatpack model was applied which fits between the magnet bore
and the vacuum tube and has a good contact with the vacuum tube surface.

In addition to the first sensor at the position of the Penning trap (see Fig.\,\ref{fig:3}),
and to the second and third at the top and bottom of the superconducting magnet bore,
a fourth one has been installed in the vicinity of the magnet for the room temperature
measurement. The central sensor is used to measure the temperature for
the control loop. The sensors are read out with a multichannel digital
multimeter (Keithley data acquisition system, model 2700), which is connected
via a GPIB (General Purpose Interface Bus) interface to a computer.

The temperature regulation system is controlled by a
LabVIEW program, which implements a PID (for Proportional, Integral, and Derivative)
regulation routine \cite{PIDRegulation}:
The temperature reading from the central temperature sensor is compared to
a set temperature value and the required heating power is supplied from a
Keithley power supply (model 2303) with a maximum output of 45\,W to a 50\,W
resistor inside a heater box.
The applied current is controlled by the PID regulation
in order to maintain the constant set bore temperature, while the
air is constantly circulated by use of two fans. 

\begin{figure}
\begin{center}
\includegraphics[scale=1.0]{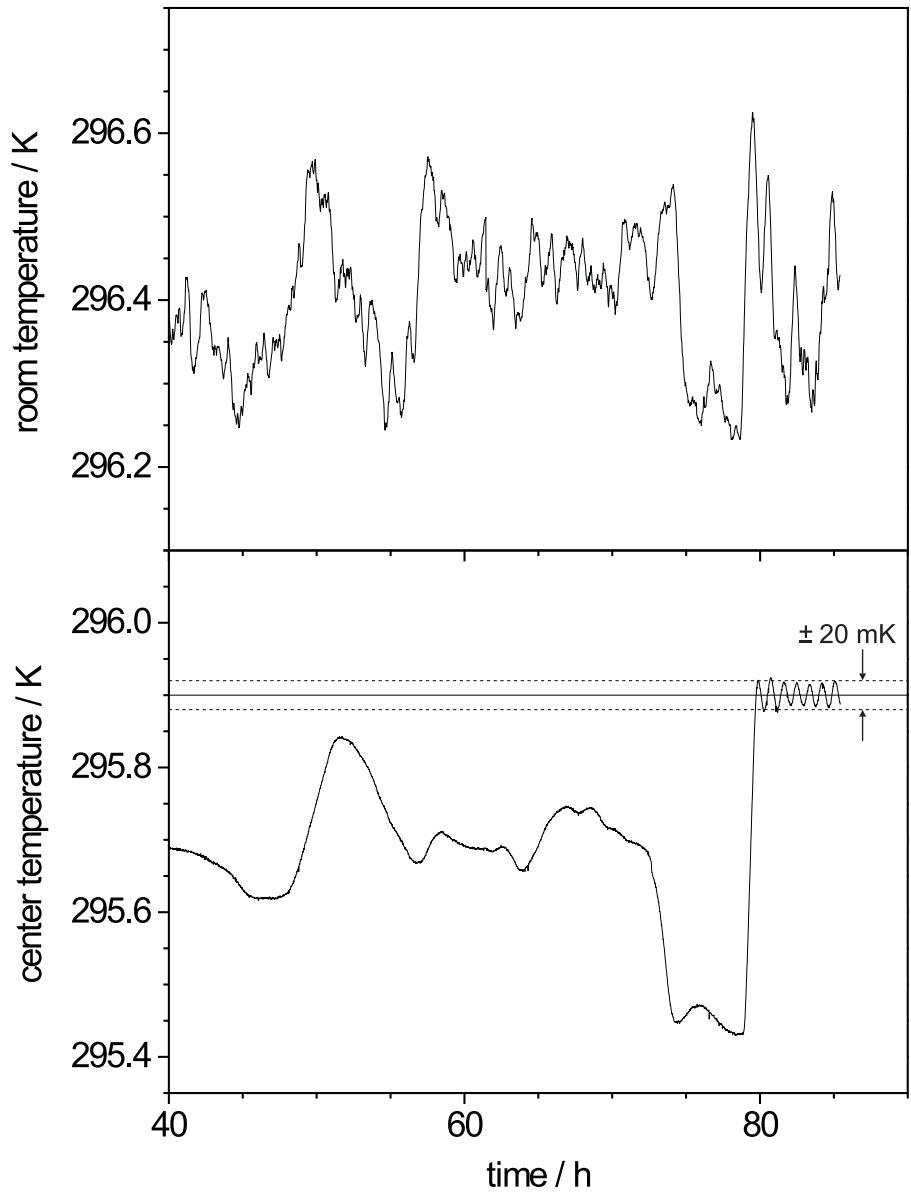}
\end{center}
\caption{\label{fig:4}
Room temperature (top) and the temperature in the vicinity of the
Penning trap vacuum tube (bottom) as a function of time
as measured by a AD590 sensor. At about $t=80$\,h
the PID regulation was turned on. The solid line shows the set temperature,
$T=295.9$\,K and the dashed lines give the range of the temperature
variations, i.e. $\pm20$\,mK.}
\end{figure}

In Fig.\,\ref{fig:4} the room temperature and the bore temperature
are shown as a function of time. Without the PID regulation the
center temperature (bottom) roughly follows the variation of the room
temperature (top). Once turned on, the center temperature is stabilized
(depending on the PID parameters), in the present case
to $\Delta T=\pm20$\,mK. 

\subsection{Helium pressure stabilization}
\begin{figure}
\begin{center}
\vspace{5cm}
\includegraphics[scale=0.8]{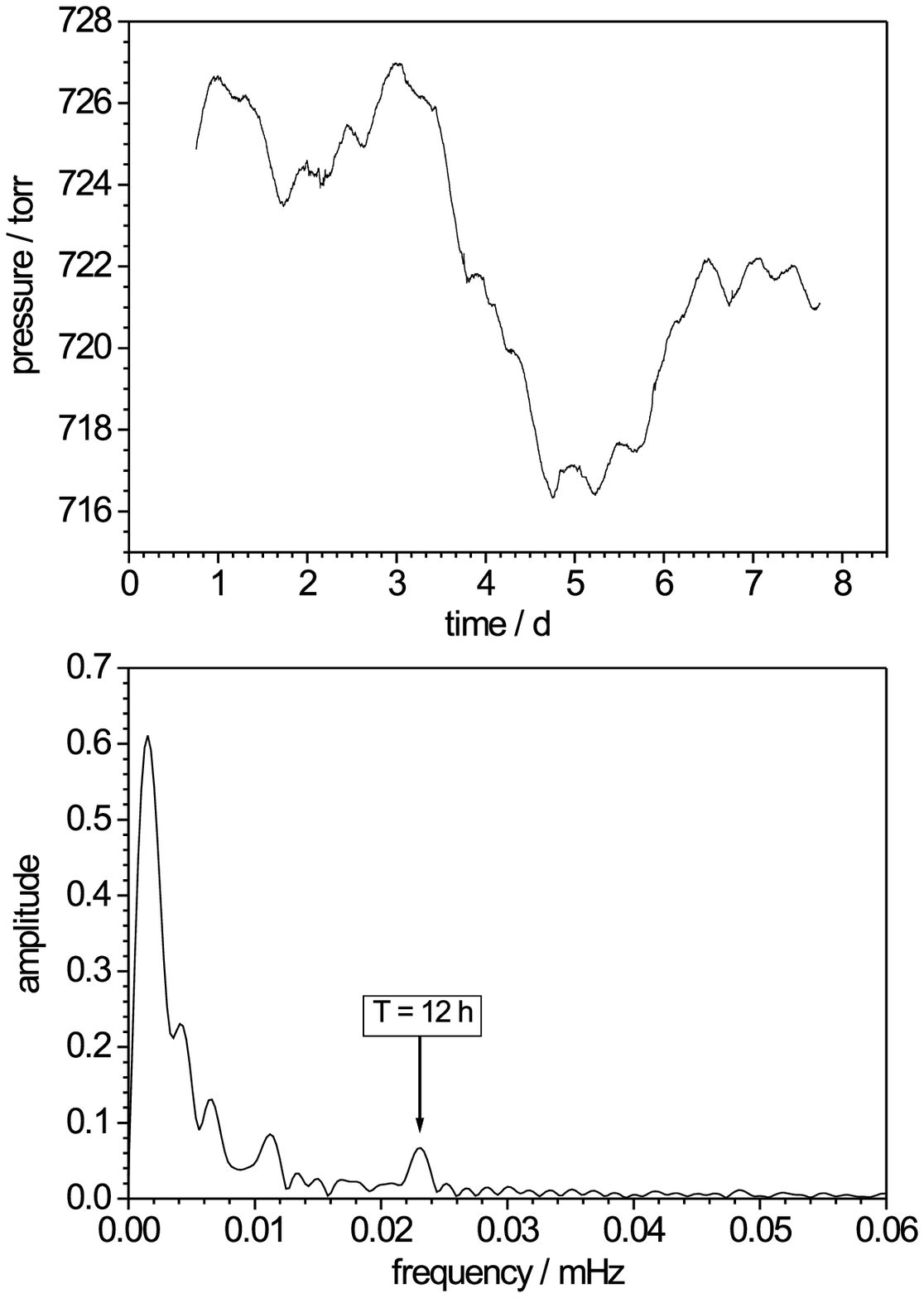}
\end{center}
\vspace{3cm}
\caption{\label{fig:5}
Top: Pressure in the helium cryostat of the superconducting magnet of the
precision Penning trap, measured in the exhaust line as a function of
time without regulation.
Bottom: Frequency spectrum of the pressure variations.}
\end{figure}

For the stabilization of the pressure in the helium
cryostat of the superconducting magnet a commercial system
from MKS has been implemented. 
Without regulation the pressure shows a behavior as plotted
in Fig.\,\ref{fig:5}\,(top). Such pressure fluctuations are
due to changes in the atmospheric pressure and changes in the gas load
of the helium-recovery line at ISOLDE.
A Fourier transform
reveals the fastest significant periodic changes with a period of $T=12$\,h,
i.e., a regulation system as presented in the following is suitable for
these slow variations.  

\begin{figure}
\begin{center}
\includegraphics[scale=1.3]{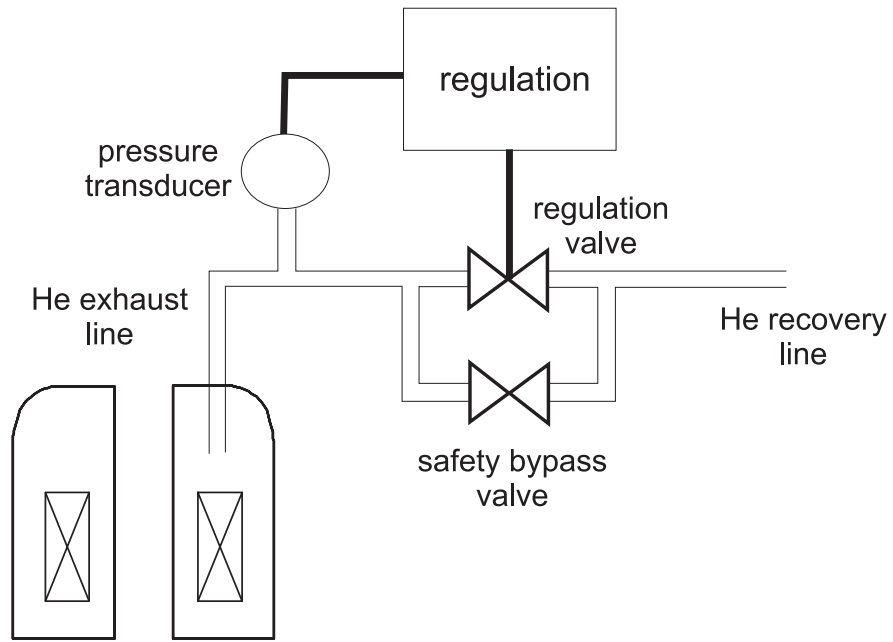}
\end{center}
\caption{\label{fig:6}
Layout of the pressure regulation system.}
\end{figure}

The layout of the regulation system is shown in Fig.\,\ref{fig:6}.
The helium exhaust line of the magnet cryostat is connected to a
regulation valve (MKS, model 248), which is controlled by a
regulation system (MKS, model 250E), that uses a PID regulation loop.
The pressure is determined by a pressure transducer (MKS, Baratron
627B), which is temperature stabilized in order to allow
reliable pressure readings independent of the room
temperature fluctuations. 

The PID controller compares the measured
pressure at the helium exhaust line with the desired set point and
adjusts the gas flow through the regulation valve in order to reach
the requested pressure. Note that the set point for the pressure
in the cryostat must be higher than the pressure in the recovery line
to prevent backstreaming of the helium into the cryostat. For security
reasons a bypass valve is opened during the refilling of the liquid helium
cryostat to allow a fast release of the larger amount of evaporated 
helium into the recovery line. 

\begin{figure}
\begin{center}
\includegraphics[scale=0.7]{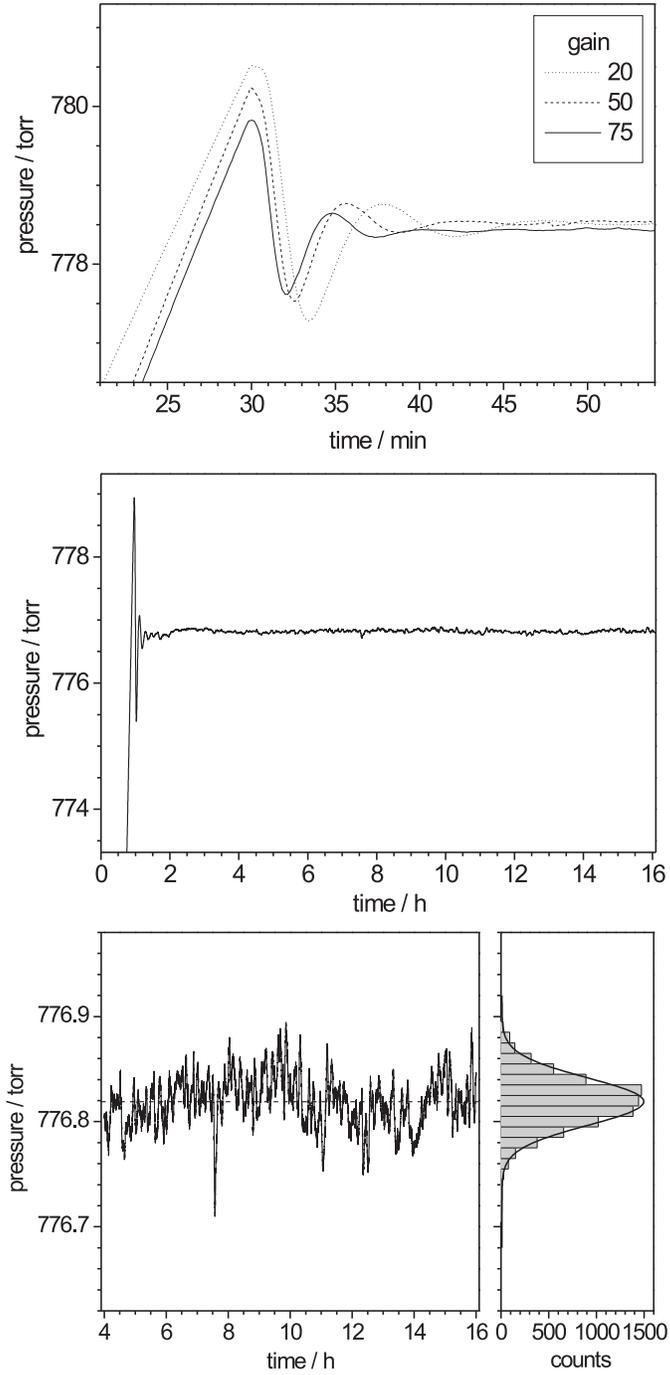}
\end{center}
\caption{\label{fig:7}
Top: Regulated pressure in the helium exhaust line as a function of time
for three different gain values. The three curves have been aligned along
the time axis with respect to the maximum value. 
 Center: Regulated pressure for a gain value of 75 as a function of time.
Bottom: Enlarged view on the pressure values, where the dashed line indicates
the mean pressure with a FWHM of 45\,mTorr of the respective distribution.}
\end{figure}

The three PID regulation parameters \cite{PIDRegulation}
are manually adjusted with two front panel potentiometers:
one for the proportional gain, which is internally combined with
the integral parameter, and one for the derivative gain
(or phase lead). As an example, the regulated pressure
is shown as a function of time in Fig.\,\ref{fig:7}\,(top) for three
different gain values for the response to the pressure difference.
With the appropriate phase lead being set according to the rapidity in
pressure changes, a gain value of 75 gives the fastest recovery of the
system in the present case.
Once set, the pressure remains constant as shown in Fig.\,\ref{fig:7}\,(middle)
where the pressure variation lies within $\Delta p(FWHM)=45$\,mtorr. 

\section{Results and discussion}

\subsection{Temperature and pressure dependence}
In order to specify the stability of the temperature and the pressure
regulation and therefore the stability of the frequency determination,
either the temperature or the pressure is deliberately changed while
the other parameter is kept fixed. The change of the cyclotron frequency
is monitored by examining stable $^{133}$Cs$^+$ ions from the alkali
reference ion source. 

\begin{figure}
\begin{center}
\includegraphics[scale=0.8]{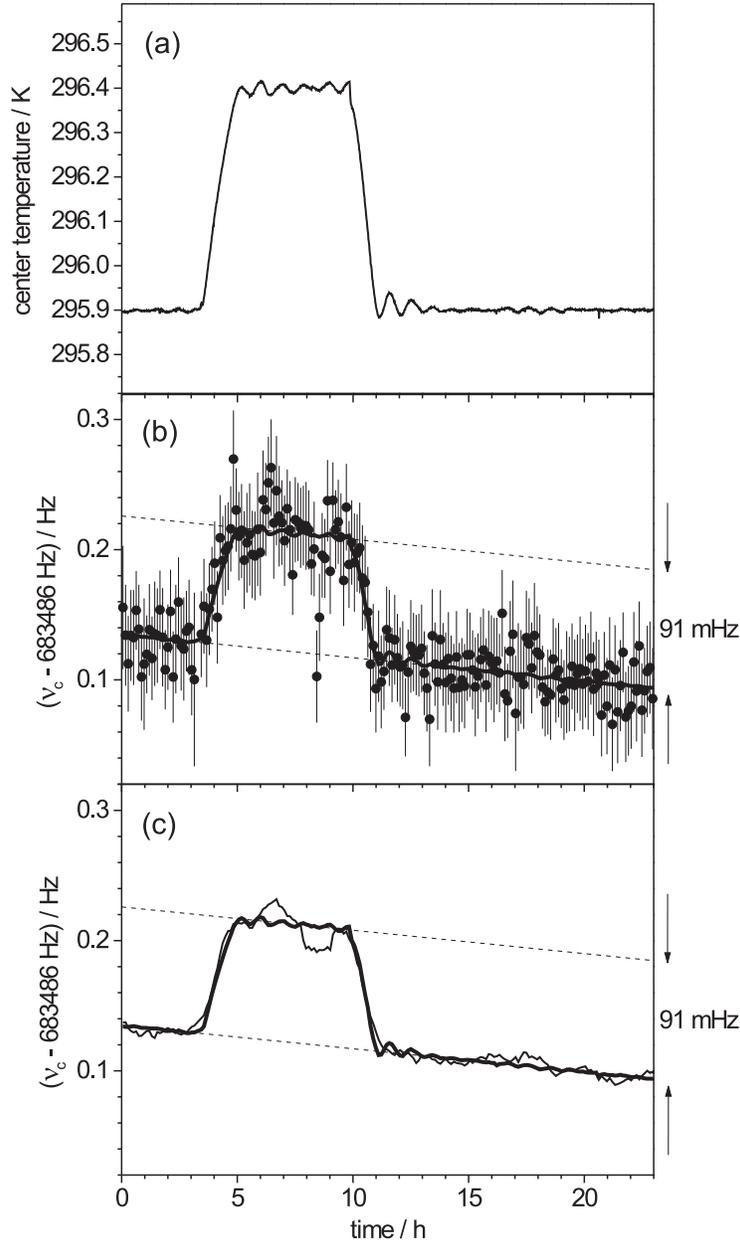}
\end{center}
\caption{\label{fig:8}
(a) Temperature in the vicinity of the Penning trap as a function of time.
At $t=3.3$\,h and $t=9.8$\,h the set value of the PID regulation has
been shifted and reset, respectively.
(b) Cyclotron frequency of $^{133}$Cs$^+$ (circles) and the
expected frequency as deduced from the temperature behavior by
fitting Eq.\,(\ref{expectedFreq}) (solid line).
The dashed lines show the linear drift as a function of time for
the two outmost temperature settings.
(c) Same as (b) for averaged frequency data (thin line) taking 10
neighboring data points into account. The drop between $t=8$\,h and 9\,h
is due to a bridge crane movement over the superconducting magnet (see also
Fig.\,\ref{fig:10}).}
\end{figure}

In Fig.\,\ref{fig:8}\,(a) the temperature in the center of the room
temperature bore of the superconducting magnet is shown as a function of
time. For the following test measurement the set temperature has
been changed deliberately from 295.9\,K to 296.4\,K, i.e., an increase by 0.5\,K, and
after about 5 hours it was reset to 295.9\,K. The resulting change of the
cyclotron frequency $\nu_c$ of $^{133}$Cs$^+$ is shown in Fig.\,\ref{fig:8}\,(b).
Each data point includes about 700 ions for a quadrupolar rf-excitation
of 900\,ms in order to keep the statistical uncertainty well below the
observed frequency shift caused by the temperature change. 

As expected a correlation of the temporal behavior of the temperature and
the cyclotron frequency is observed. Taking the linear temporal drift
of the magnetic field into account, the expected cyclotron frequency
at a temperature $T$ in the room temperature bore
for a fixed and stabilized helium reservoir pressure can be expressed by
\begin{equation}
\label{expectedFreq}
\nu_c(T,t)=a(T-T_0)-bt+c,
\end{equation}
which is the analogue expression to the one in Eq.\,(\ref{Eq:resistance}).
From a $\chi^2$-minimization with a fixed value $T_0=295.9$\,K the
parameters $a=0.182(13)$\,HzK$^{-1}$, $b=0.0018(4)$\,Hzh$^{-1}$, and
$c=683486.135(6)$\,Hz, are deduced for the $^{133}$Cs data.
The field drift is thus
\begin{equation}
\frac{1}{B}\frac{dB}{dt}=-\frac{b}{c}=(-2.6\pm0.6)\times10^{-9}\,\mbox{h}^{-1}
\end{equation} 
in agreement to the previously determined value.

For a 0.5\,K temperature change, the observed change of the $^{133}$Cs$^+$
cyclotron frequency is 91\,mHz (see Fig.\,\ref{fig:8}\,(b)).
Assuming a linear dependence, this yields a temperature coefficient of
0.182\,mHz/mK and for the planned stabilization of the temperature to
$\pm5$\,mK this results in a relative frequency change of
$2.7\times10^{-9}$.

\begin{figure}
\begin{center}
\includegraphics[scale=0.8]{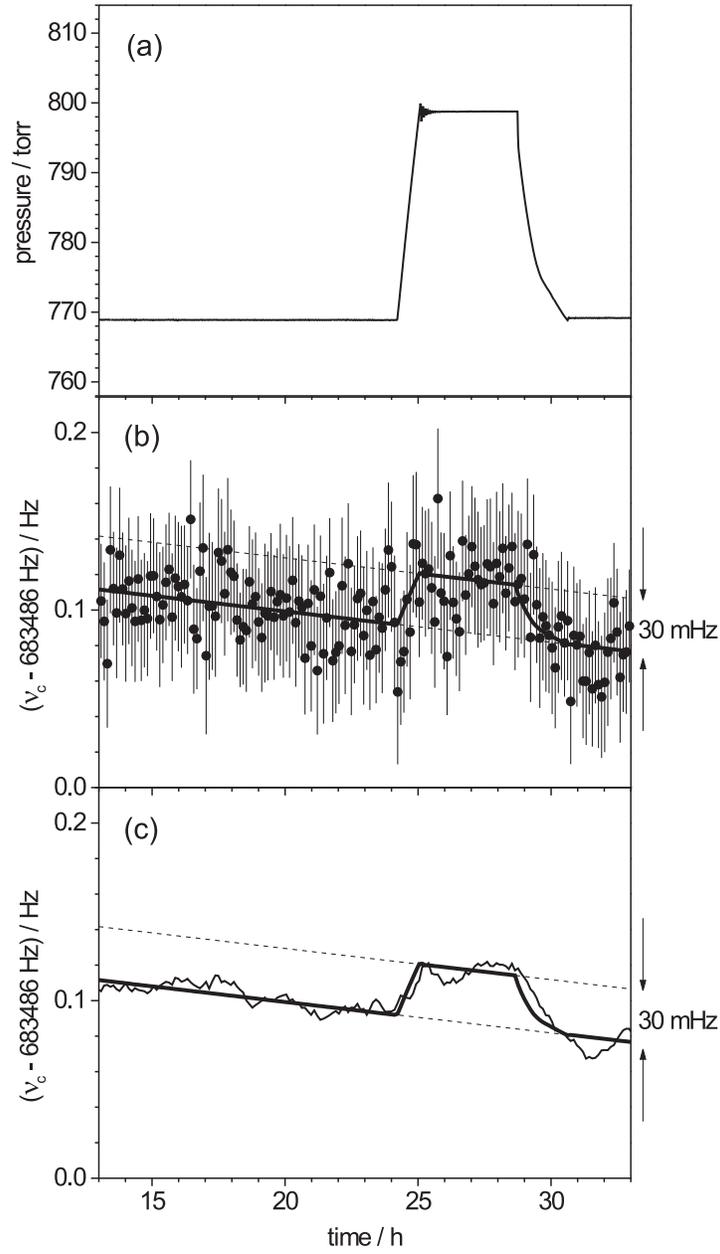}
\end{center}
\caption{\label{fig:9}
(a) Pressure in the helium exhaust line as a function of time.
At $t=24.2$\,h and $t=28.7$\,h the set value of the pressure regulation has
been shifted by 30\,torr and reset, respectively.
(b) Cyclotron frequency of $^{133}$Cs$^+$ (circles) and the
expected frequency as deduced from the pressure behavior by
fitting an analogue expression of Eq.\,(\ref{expectedFreq}) (solid line).
The dashed lines show the linear drift as a function of time for
the two outmost pressure settings.
(c) Same as (b) for averaged frequency data (thin line) taking 10
neighboring data points into account.}
\end{figure}

A similar measurement has been performed for the change of the
pressure in the liquid helium cryostat. The result is shown
in Fig.\,\ref{fig:9}. The set pressure of the regulation system
has been changed by 30 torr.
A fit to the data points of a linear relation in analogy
to Eq.\,(\ref{expectedFreq}), $\nu_c(p,t)=a(p-p_0)-bt+c$,
yields a frequency change of 1\,mHz/torr. Thus, for a
stabilization of the pressure to $\pm0.1$\,torr a relative frequency
change of $2.9\times10^{-10}$ can be achieved.
Note that in contrast to the averaged experimental data in case of a
temperature change (Fig.\,\ref{fig:8}\,(c), thin line), the averaged data
for a pressure change (Fig.\,\ref{fig:9}\,(c), thin line) shows a slight
delay as compared to the expected behavior (thick line). Nevertheless, this
might be an artefact and has no significant influence.

\subsection{Other effects}
In addition to magnetic field variations that are correlated to
temperature and pressure changes, the magnetic field strength can
also be influenced by ferromagnetic metallic objects in the vicinity
of the superconducting magnet,
which can distort the magnetic field. In the case of ISOLTRAP
the beam of a bridge crane mounted at the ceiling of the experimental hall
of ISOLDE can cause large frequency shifts if it is placed above the
superconducting magnet of the precision Penning trap. 

\begin{figure}
\begin{center}
\includegraphics[scale=1.0]{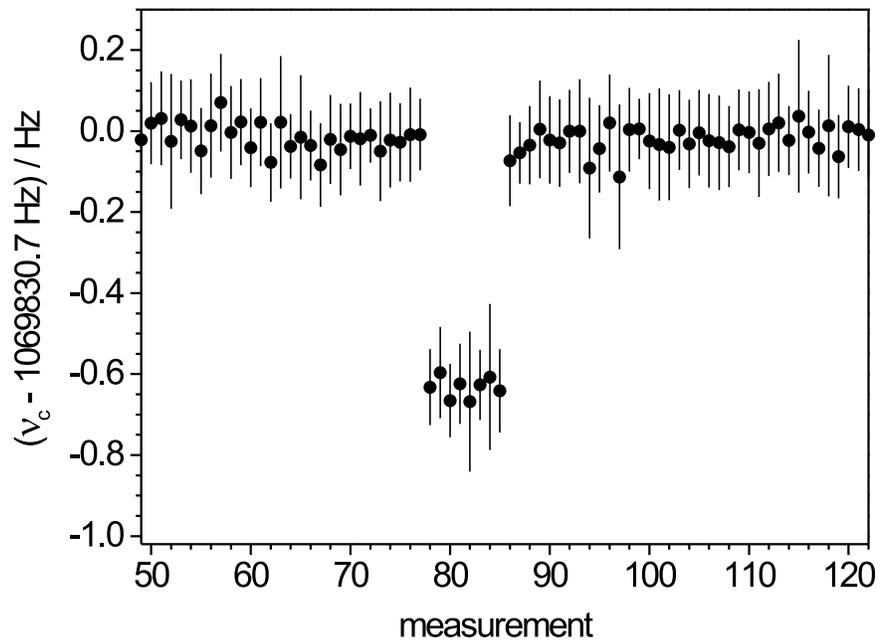}
\end{center}
\caption{\label{fig:10}
Cyclotron frequency of $^{85}$Rb$^+$ for subsequent measurements
(total data collection time about 1 minute in each case)
when a bridge crane was moved close to the superconducting magnet
at the measurement time of data points $78-86$. Before and after
the crane was placed in far distance.}
\end{figure}

In Fig.\,\ref{fig:10} a consecutive series of measurements of the cyclotron
frequency of $^{85}$Rb$^+$ ions is shown. When the steel beam of the bridge crane
was moved over the magnet, the cyclotron frequency dropped about 600\,mHz due
to the distortion of the magnetic field. This corresponds to
a relative change of $5.6\times10^{-7}$. As soon as the bridge crane is moved away from
the magnet, the cyclotron frequency returns to its previous value. 
There is no attempt to compensate for this change, rather it is avoided to move
the beam of the bridge crane over the magnet during data collection.

\subsection{Stabilized frequency measurement}
\begin{figure}
\begin{center}
\includegraphics[scale=1.2]{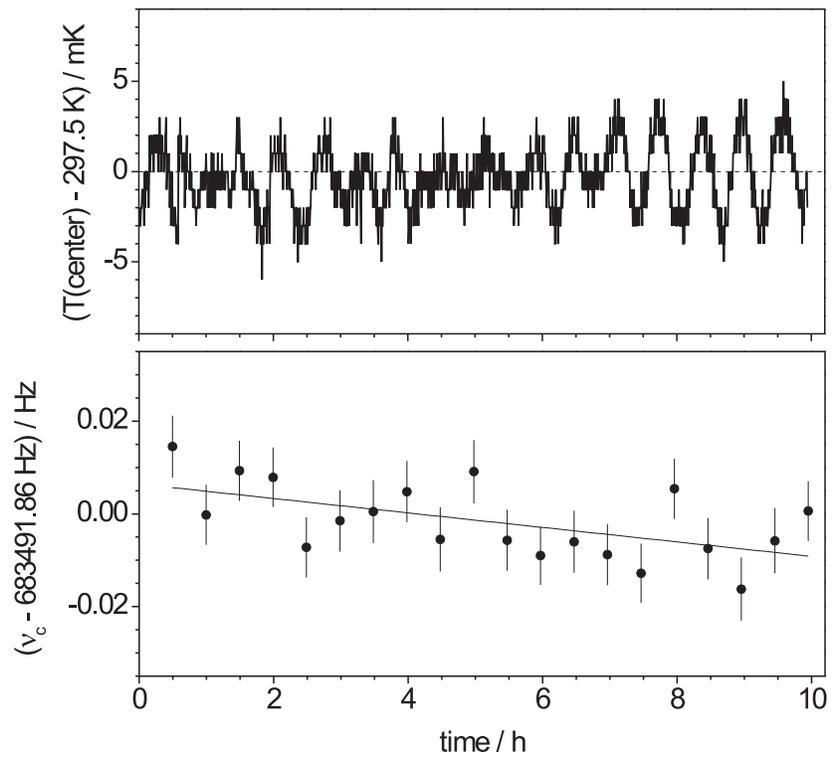}
\end{center}
\caption{\label{fig:11}
Center temperature (top) and cyclotron
frequency of $^{133}$Cs$^+$ (bottom) as a function of time.
The solid line is a linear fit to the frequency data points.}
\end{figure}

Fig.\,\ref{fig:11} shows a measurement of the cyclotron
frequency of $^{133}$Cs$^+$ for which a temperature stabilization
to $\pm5$\,mK has been achieved. Within the statistical uncertainty the
frequency data points show only the linear temporal drift due to the
residual resistance of the superconducting coils. Note that the room
temperature fluctuations have been well below 500mK during the data taking.
Since the optimal PID paramaters have not been found yet, the general
performance of the stabilization system still needs to be investigated and
only the influence of the temperature variations on the frequency measurements
can be deduced. 

\begin{figure}
\begin{center}
\includegraphics[scale=1.0]{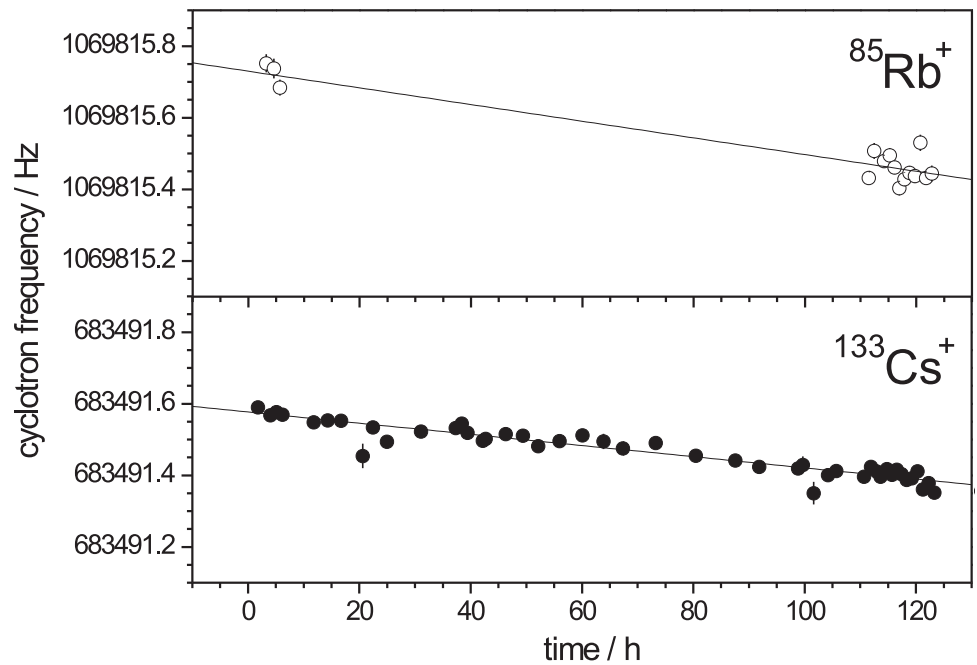}
\end{center}
\caption{\label{fig:12}
Cyclotron frequency of $^{85}$Rb$^+$ (top) and $^{133}$Cs$^+$
(bottom) as a function of time. The solid lines are linear fits 
to the data points.}
\end{figure}

Figure\,\ref{fig:12} shows cyclotron frequency data from $^{85}$Rb$^+$ and $^{133}$Cs$^+$
that were measured as references during a beam time for the determination of the mass of
neutron-rich Sn nuclides. The masses of these stable nuclides are known with a relative
uncertainty $\delta m/m=1.4\times10^{-10}$ and $1.8\times10^{-10}$, respectively \cite{Bradley}.
The current limit for the mass determination at ISOLTRAP
is $\delta m/m=8\times10^{-9}$ \cite{KellerbauerEPJD}, i.e. any fluctuation or systematic shifts
can be probed with the two reference nuclides.
The lines in Fig.\,\ref{fig:12} are weighted linear fits
to the data points. The B-field drifts for $^{85}$Rb$^+$ and $^{133}$Cs$^+$
are $(-2.2\pm0.2)\times10^{-9}$\,h$^{-1}$ and $(-2.1\pm0.1)\times10^{-9}$\,h$^{-1}$,
respectively. The fits also yield the offsets (at $t=0$) $\nu_c=1\,069\,815.730(15)$\,Hz
and 683\,491.576\,9(37)\,Hz, respectively, and therefore a frequency ratio 0.638\,887\,200\,6(96),
where the uncertainty is mainly due to the low statistics of the $^{85}$Rb$^+$ data.
From the literature values a frequency ratio 0.638\,887\,196\,90(15) is expected, i.e.
the measured value agrees within the uncertainties.

\begin{figure}
\begin{center}
\includegraphics[scale=1.0]{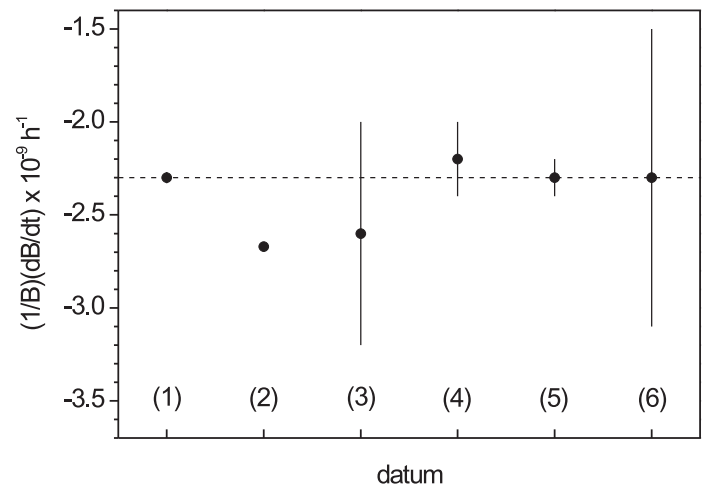}
\end{center}
\caption{\label{fig:13}
Magnetic field drift $(1/B)(dB/dt)$ as given in Ref.\,\cite{KellerbauerEPJD} (value (1), corrected),
and as deduced from the data shown in: (2) Fig.\,\ref{fig:2}, (3) Fig.\,\ref{fig:8}, (4) Fig.\,\ref{fig:11},
(5) Fig.\,\ref{fig:12} for $^{85}$Rb$^+$, and (6) Fig.\,\ref{fig:13} for $^{133}$Cs$^+$. The dashed line
shows the corrected value of \cite{KellerbauerEPJD} as a reference.}
\end{figure}

Finally, the linear drift of the magnetic field as deduced from the various investigations in
the present work can be compared to the value given in \cite{KellerbauerEPJD}. A summary is shown
in Fig.\,\ref{fig:13}. There is a good agreement except in the case of the data presented in
Fig.\,\ref{fig:2}, for which a significant shift is observed. Possibly the uncertainty is underestimated.
In addition, the linear drift must not necessarily have a constant value within several years. 
Nevertheless, a value of $-2.30\times10^{-9}$\,h$^{-1}$ seems to be a reasonable
estimate for the linear drift. 

\section{Summary and outlook}
The magnetic field in the vicinity of the precision Penning trap
of ISOLTRAP has been stabilized with respect to the temperature
in the room-temperature bore of the magnet as well as to the
pressure in the helium cryostat. A strong correlation of
the cyclotron frequency to the temperature of the vacuum tube
around the Penning trap has been observed. With a
stabilization down to $\pm5$\,mK only a negligible linear decrease of the
cyclotron frequency due to the finite resistance of the
superconducting coils was observed within the statistical limits.
The parameters of the PID temperature regulation are currently under
investigation to further decrease the remaining small oscillations of the
regulated temperaturea and to obtain a more robust system which is
capable of stabilizing the center temperature to $\pm5$\,mK also
for larger room-temperature fluctuations.
In addition, a stabilization
of the pressure in the liquid helium reservoir of the magnet to
less than $\pm50$\,mtorr was achieved. With these stabilization systems
fluctuations of the magnetic field have been reduced
by a factor 4 below the current limit of accuracy of ISOLTRAP. The
improvement with respect to the ISOLTRAP mass measurements will be 
investigated soon.

\section*{Acknowledgements}
This work was supported by
the German Ministry for Education and Research (BMBF) under
contracts 06GF151 and 06GF181I,
the European Commission under contracts
HPMT-CT-2000-00197 (Marie Curie Fellowship)
and RII3-CT-2004-506065 (EURONS/TRAPSPEC),
and by the Helmholtz association of national research centres (HGF)
under contract VH-NG-037.
We also acknowledge stimulating discussions with T. Fritioff from
the University of Stockholm, Sweden.

%%%%%%%%%%%%%%%%%%%%%%%%%%%%%%%%%%%%%%%%%%%%%%%%%%%%%%%%%%%%%%%%%%%%%%%%%%%%
%
% figures
%
%%%%%%%%%%%%%%%%%%%%%%%%%%%%%%%%%%%%%%%%%%%%%%%%%%%%%%%%%%%%%%%%%%%%%%%%%%%%
\clearpage
%Fig. 1

\clearpage
%Fig. 2

\clearpage
%Fig. 3

\clearpage
%Fig. 4

\clearpage
%Fig. 5
\vspace{11cm}

\clearpage
%Fig. 6

\clearpage
%Fig. 7

\clearpage
%Fig. 8

\clearpage
%Fig. 9

\clearpage
%Fig. 10

\clearpage
%Fig. 11

\clearpage
%Fig. 12

\clearpage
%Fig. 13

\end{document}